# Chiral Surface Plasmon Polaritons on Metallic Nanowires

Shunping Zhang,[1] Hong Wei,[1] Kui Bao,[2] Ulf Håkanson,[1,4] Naomi J. Halas,[1,2,3] Peter Nordlander,[1,2] and Hongxing Xu[1,4,*]

[1]*Beijing National Laboratory for Condensed Matter Physics and Institute of Physics, Chinese Academy of Sciences, Box 603-146, Beijing 100190, China*
[2]*Department of Physics and Astronomy, Laboratory for Nanophotonics, Rice University, 6100 Main Street, MS-366, Houston, TX 77005, USA*
[3]*Department of Electrical and Computer Engineering, Rice University, 6100 Main Street, MS-366, Houston, TX 77005, USA*
[4]*Division of Solid State Physics/The Nanometer Structure Consortium (nmC@LU), Lund University, Box 118, S-22100, Lund, Sweden*

**Abstract**

Chiral surface plasmon polaritons (SPPs) can be generated by linearly polarized light incident at the end of a nanowire, exciting a coherent superposition of three specific nanowire waveguide modes. Images of chiral SPPs on individual nanowires obtained from quantum dot fluorescence excited by the SPP evanescent field reveal the chirality predicted in our theoretical model. The handedness and spatial extent of the helical periods of the chiral SPPs depend on the input polarization angle and nanowire diameter as well as the dielectric environment. Chirality is preserved in the free-space output wave, making a metallic nanowire a broad bandwidth subwavelength source of circular polarized photons.

An object is chiral when it is not identical to its mirror image. An electromagnetic wave can be chiral when it carries a nonzero angular momentum: an example of this is a circularly polarized plane wave. The interaction of circularly polarized electromagnetic fields with chiral objects, such as molecules, is a topic of fundamental interest, which also has important applications in fields such as molecular spectroscopy and structural biology [1-4]. Recent advances in our understanding of how to shape electromagnetic fields at nanoscale dimensions have led to the concept of tailoring the optical chirality to match molecular dimensions, and to the design and demonstration of chiral nanostructures [5-8]. Metallic structures play an important role in addressing this challenge, since they support surface plasmon polaritons (SPPs), which are collective excitations of their conduction electrons. SPPs enable localized and highly spatially structured electromagnetic fields that can be scaled down to nanometer dimensions [9, 10].

Single crystalline metallic nanowires have recently attracted significant interest as subwavelength SPP waveguides, analogs of optical fiber waveguides but with nanometer-scale cross sections and micron-scale propagation lengths. The strong coupling between proximal quantum emitters and nanowire SPPs, facilitated by the small mode volume, can enable nanowires to serve as quantum information transmission lines [11-14]. Nanowires can function as low-Q Fabry-Pérot resonators [15, 16] and as unidirectional subwavelength light sources, by exciting SPP modes with light incident at a nanowire terminus [17, 18]. Active control of the phase or polarization state of the incident excitation source can generate coherent superpositions of nanowire SPPs. This approach has given rise to a series of nanowire-based plasmonic devices, such as routers, modulators, even all-optical Boolean logic gates that can perform simple computational operations [19-21].

In this letter, we show that chiral SPPs can be generated on metallic nanowires by exciting a superposition of nanowire SPP modes. We find that the coherent excitation of three specific plasmon modes - $HE_1$ and $HE_{-1}$ modes with a $\pi/2$ phase delay and the fundamental $TM_0$ mode - gives rise to chiral plasmons. This complex excitation is easily achieved by illuminating a nanowire at one end with linearly polarized light at nominally 45 degrees with respect to the nanowire axis. The helical local field distributions of the chiral plasmons are confirmed by quantum dot imaging of individual nanowires. Upon propagation to the distal end of the nanowire, the chirality of the waveguide mode is preserved in the emitted photon. This property makes the nanowire a subwavelength λ/4 converter of the input light. Chiral nanowire plasmons and their circularly polarized emitted photons may be useful tools for tailoring the interaction between enantiomeric molecules and superchiral electromagnetic fields, a topic of keen current interest [2, 3].

Due to the momentum and energy mismatch between photons and plasmons, optical excitation of SPPs in metallic nanowires requires local symmetry breaking. [16, 22, 23]. This can be achieved quite simply by optical excitation at the nanowire terminus. As shown in Fig. 1(a), a cylindrical nanowire with permittivity $\varepsilon^M$, radius $R$ and length $L$ is aligned parallel to the x-axis and terminated by two flat facets so that the input end is centered at the origin and the distal end at ($L$, 0, 0). A paraxial Gaussian beam (waist $w_0 = 1.0\ \mu m$) polarized at an angle θ to the wire axis, is directed normally (along the z-axis) onto the input end. The nanowire is embedded in a homogenous background with permittivity $\varepsilon^D$. Optical constants for Ag are interpolated from Ref. [24]. All materials in the system are assumed to be nonmagnetic and modeled with unity permeability.

For an infinitely long metallic nanowire, the surface plasmon modes can be modeled analytically [13, 25, 26]. In cylindrical coordinates, the electric field takes the form $\mathbf{E}^j(\mathbf{r}) = \sum_m a_m^j \mathbf{E}_m^j(k_{m,\perp}^j \rho) e^{im\phi} e^{ik_{m,\|}x}$, where the superscript $j = $ D, M represents the region outside (dielectric) and inside (metal) the wire, respectively, and $m$ is the azimuthal quantum number. $a_m^j$ and $k_{m,\|}$ are the amplitude and the propagating constant of the $m^{\text{th}}$ mode, respectively, and $(\rho, \phi, x)$ are the cylindrical coordinates. $k_{m,\|}$ is related to the transverse wave vector $k_{m,\perp}^j$ by $k_{m,\parallel}^2 + k_{m,\perp}^{j\,2} = \varepsilon^j k_0^2$, where $\varepsilon^j$ is the dielectric constant in region $j$ and $k_0$ is the wavenumber in vacuum. The TM$_0$ ($m = 0$) mode is a no-cutoff, axially symmetric mode resulting from electrons oscillating parallel to the wire axis. The two degenerate first order modes, HE$_{-1}$ ($m = -1$) and HE$_1$ ($m = 1$), correspond to charge oscillations in the vertical and horizontal plane, respectively. These two modes exhibit an exponentially growing increase of their mode volume in the thin wire limit [13].

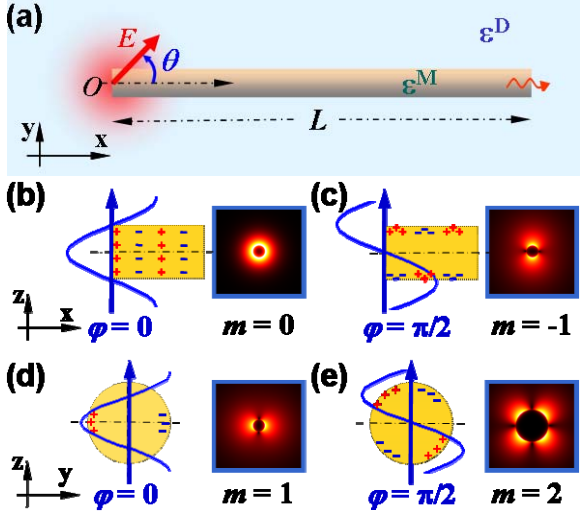

**FIG. 1** (a) Schematic illustration of the nanowire in homogeneous background. The incident light is focused on the left nanowire end. (b) Excitation of the $m = 0$ (TM$_0$) and (c) $m = -1$ (HE$_{-1}$) mode at an incident phase $\varphi = 0$ and $\varphi = \pi/2$, respectively, for incident electric field polarized parallel to the wire axis. (d) Excitation of the $m = 1$ (HE$_1$) and (e) $m = -2$ (HE$_2$) mode at $\varphi = 0$ and $\varphi = \pi/2$, respectively, for perpendicular incident polarization. The electric field $|\mathbf{E}|$ profile of each excited SPP mode is shown on the right in panels (b-e).

The nanowire system illustrated in Fig. 1 was modeled using the finite element method (FEM) method (Comsol Multiphysics 3.5a, RF Module). For a finite length wire, local symmetry-breaking at the wire terminus is sufficient to allow the SPPs to be excited by incident light via near-field coupling. Generally speaking, the excitation of a particular SPP requires an overlap between the mode profiles of the SPP and the excitation source. When retardation effects are significant, such as when the size of a nanowire is large compared to the incident wavelength, several plasmonic modes can overlap the excitation source in this geometry [27].

We now consider the excitation of SPPs in a thick nanowire (one with a diameter larger than the quasistatic limit) using the excitation geometry depicted in Fig. 1(a). The incident excitation is described by a Gaussian beam with an instantaneous electric field of the form $\mathbf{E}_{inc}(\mathbf{r}) = \mathbf{E}_0(\mathbf{r}) \cdot e^{-i\varphi}$, where $\varphi = \omega t$ is the incident phase and $\mathbf{E}_0(\mathbf{r})$ is the mode profile of the incident light. Excitation of specific SPP modes is accomplished by aligning the incident polarization either parallel or perpendicular to the wire axis and by selecting the appropriate phase of the optical field at the input (Fig. 1(b-e)). For $\varphi = 0$, selective excitation of the $TM_0$ or $HE_1$ mode is achieved by aligning the incident polarization parallel or perpendicular to the wire axis, as illustrated in Fig. 1(b) and Fig. 1(d), respectively. With a π/2 phase shift of $\varphi$, the local field profile across the nanowire terminus changes in symmetry creating an electric field that is anti-symmetric across the $z = 0$ plane. The excitation profile now matches the mode distribution of $HE_{-1}$ (Fig. 1(c)) or $HE_2$ (Fig. 1(e)) for parallel or perpendicular incident polarization, respectively. Therefore for an arbitrary input phase $\varphi$, both $TM_0$ and $HE_{-1}$ modes are excited for parallel incident polarization, and both $HE_1$ and $HE_2$ modes are excited for perpendicular polarization. Experimentally we should see the cycle-averaged superposition of all excited modes, with amplitudes dependent on the ratio of nanowire diameter to the wavelength of incident light.

For an arbitrary incident polarization angle $0° < \theta < 90°$, the three lowest modes ($|m| \leq 1$) are excited simultaneously, but in a manner such that the $HE_{-1}$ mode has a constant phase delay $\Delta\Phi = \pi/2$ with respect to the $HE_1$ mode. A full range of elliptical SPPs can be generated in this manner. The coherent interference of these two SPP waves of equal amplitude results in a circularly polarized guided wave. We refer to such a wave as a chiral SPP, characterized by its circular angular momentum $\sigma_\pm$, in direct analogy with photons. The contribution of the simultaneously excited $TM_0$ mode is to stretch the chiral SPP into a helical wave, with a spiral near-field pattern. Figure 2(a) shows the surface charge distribution of a chiral SPP on a 120 nm diameter nanowire locally excited by a Gaussian beam, for incident phase $\varphi = 0.683\pi$ and polarization angle $\theta = 45°$. A helical propagating SPP resulting from the superposition of the three lowest SPP modes is clearly observed. The time-averaged power flow at various cross sectional positions of the nanowire, shown in Fig. 2(b), further confirms this helical behavior. As indicated by the white arrow, the electromagnetic energy revolves around the metal-dielectric interface as the SPP propagates along the nanowire. In the example shown here, the period of the helix ($\Lambda$) is about 1.83 μm.

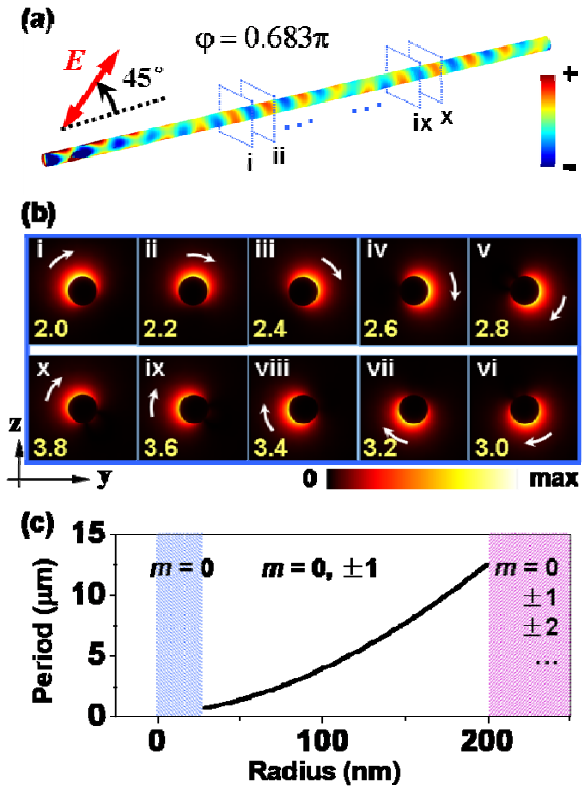

**FIG. 2** Chiral SPPs. (a) Normalized surface charge density plot on a silver nanowire. The maximal value of surface charge was truncated at the incident end to better show the plasmon modes on the wire. (b) Time-averaged power flow in the yz-plane at different positions along the nanowire, where $x = 2.0$ μm to $3.8$ μm (i-x) in steps of 0.2 μm, indicated by the blue frames in (a). The white arrows highlight the rotation of electromagnetic energy as a function of position along metal nanowire, showing right-handed chiral SPPs. (c) Periods of the plasmon helix, $\Lambda$, as a function of nanowire radius. The blue region denotes the single mode-dominant regime and the magenta region denotes the multi-mode regime. In (a-c), the simulated nanowire ($R = 60$ nm) is embedded in oil ($\varepsilon^D = 2.25$) and excited by a linear polarized Gaussian beam ($\lambda_0 = 632.8$ nm). The length of the nanowire is $L = 5.0$ μm and the incident polarization angle is $\theta = 45°$ in (a) and (b).

The handedness of the chiral SPP is determined by the phase delay between the $HE_1$ and $HE_{-1}$ modes. An advanced phase for the $HE_1$ mode relative to the $HE_{-1}$ mode corresponds to a clockwise rotation of the collective electron motion as viewed in the direction of propagation, resulting in a right-handed SPP $|\sigma_+ = 1\rangle$. The converse is true for a left-handed SPP $|\sigma_- = -1\rangle$ which can be excited by rotating the incident polarization from $\theta$ to $-\theta$. The relative amplitudes of the $HE_1$ and $HE_{-1}$ modes are proportional to the perpendicular and parallel components of incident beam with respect to the nanowire axis, hence, a direct tailoring of circularly or elliptically polarized SPPs can be achieved simply by varying the polarization angle $\theta$ of the incident excitation field. Therefore, a controllable spatial distribution of the surface plasmon energy can be achieved. For parallel or perpendicular excitation, either one or two of the three constituent modes will not be excited, resulting in a disappearance of the helical characteristic of the guided wave. However, since the excitation of more than one plasmon mode is still likely, a spatially dependent interference, or "beat", between the remaining excited plasmons may still be observable.

The period of a chiral SPP is proportional to the inverse of the difference between the propagation constants of the $TM_0$ and $HE_1$ (or $TM_0$ and $HE_{-1}$) mode:

$\Lambda = 2\pi(\text{Re}(k_{0,\parallel} - k_{1,\parallel}))^{-1}$. By numerically solving the transcendental equation [13], the period of a chiral SPP in a Ag nanowire as a function of wire radius is obtained (Fig. 2(c)). Here we observe that chiral SPPs can be launched on a nanowire for only a finite range of nanowire radii. Since phase retardation is critical to the formation of a chiral SPP, the diameter of the nanowire must be comparable to the excitation wavelength $R \sim \lambda_0/\sqrt{\varepsilon^D}$ to support chiral SPPs. In this regime, for increasing nanowire radius, the propagation constant of the $TM_0$ mode decreases and that of the $HE_1$ mode increases [26], yielding a larger helical period. For very thin nanowires ($R \ll \lambda_0/\sqrt{\varepsilon^D}$), the helices (and also the "beats") disappear because the structure will only support $m = 0$ modes. For larger wires, the excitation of higher order modes ($|m| \geq 2$) results in complicated field distributions that countermand the chirality of SPPs. The chiral SPPs we report here will most likely appear between these two regimes. Besides the diameter of the nanowire, its composition, embedding medium and excitation wavelength will all affect the excitation of chiral SPPs.

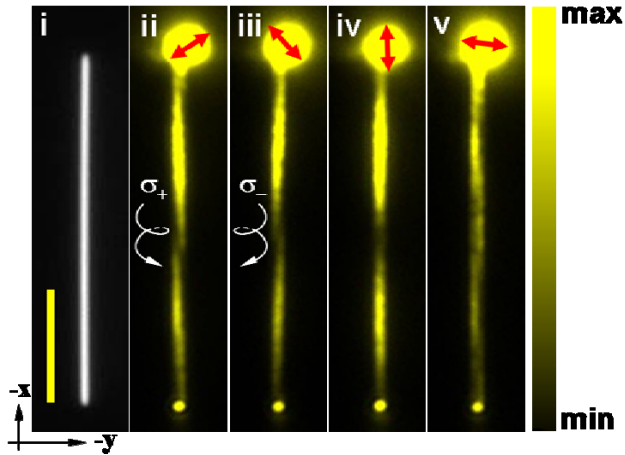

**FIG 3.** (i) Optical image of a silver nanowire. The scale bar is 5 μm. (ii, iii) Quantum dot images of (ii) a right-handed and (iii) left-handed SPPs, respectively. The white helical arrows highlight the handedness of the plasmon helix. (iii, iv) SPPs launched with incident polarization (iii) parallel and (iv) perpendicular to the nanowire axis

show suppression of the plasmon helix. All images (ii-v) are obtained by quantum dot imaging, with He-Ne laser excitation (632.8 nm).

Chiral SPPs were directly observed experimentally by means of quantum dot-based fluorescence imaging of the nanowire evanescent field. The method used here is quite similar to that reported previously [21]. Ag nanowires (about 300 nm in diameter) were deposited on a glass substrate, followed by a 10 nm alumina oxide coating to prevent quenching of the subsequently deposited quantum dots. Here it was found that a homogeneous dielectric embedding medium for the nanowire was critical for the observation of chiral SPPs. This was achieved using index-matching oil to preserve the cylindrical symmetry of the system. A set of fluorescent images obtained in this manner are shown in Figure 3. Here we see that the helical near field is clearly resolved when the incident polarization is oriented at $45°$ (ii) or $-45°$ (iii) to the wire axis. The handedness of the helix is reversed by rotating the polarization from $45°$ to $-45°$. The helical field pattern disappears when the polarization is parallel (iv) or perpendicular (v) to the nanowire axis, also in agreement with our theoretical analysis.

A direct consequence of the generation of chiral SPPs is the chirality of the emission light at the nanowire output. To evaluate the chiral SPP output field as a circularly polarized light source, we calculate the degree of circular polarization C and the Figure of Merit (FoM) $f = I \times C^2$ for this structure [28]. The definitions of these quantities have been slightly modified to take the longitudinal field components into account, namely [28]:

$$C = \frac{2\langle E_y(t)E_z(t)\sin(\delta_y - \delta_z)\rangle}{\langle E_x^2(t)\rangle + \langle E_y^2(t)\rangle + \langle E_z^2(t)\rangle}, \quad (1)$$

where $\langle\ \rangle$ denotes time average, and $\delta_y - \delta_z$ is the phase difference between the two transverse electric field components $E_y$ and $E_z$. The relative intensity is $I = |\mathbf{E}(\mathbf{r})|^2 / |\mathbf{E}_0(0)|^2$, where $\mathbf{E}_0(0)$ represents the incident electric field at the origin. Figure 4(a) shows a spatial map of C in a vertical plane 200 nm beyond the output end of the nanowire. As expected, the emitted photon preserves the chirality of the SPP. Despite the presence of a nonzero longitudinal field component, a high degree of circular polarization ($C \sim 0.90$) is obtained. The FoM (*f*), shown in Fig. 4(b), further confirms the high degree of circular polarization of the outgoing optical wave. The ellipticity of the polarization state of the emitted photon can be tuned in a continuous fashion by rotating the incident polarization angle $\theta$ (Fig. 4(c)). Higher C appears around $\theta = 45°$, when the $HE_1$ and $HE_{-1}$ modes have nearly equal amplitudes. Unlike conventional quarter-wave plates or resonance-based subdiffraction circularly polarized sources [28], metallic nanowires support a continuum of SPPs, making this geometry ideal as a broadband source of circularly polarized light. Figure 4(d) shows that photons emitted from the nanowire maintain a high degree of circular polarization ($C > 0.50$) over a wide wavelength window, spanning most of the visible region of the spectrum The modulations appearing in $C(\lambda_0)$ (red) are due to etaloning in the nanowire and also appear in the transmission spectrum $I(\lambda_0)$ at the nanowire output (green).

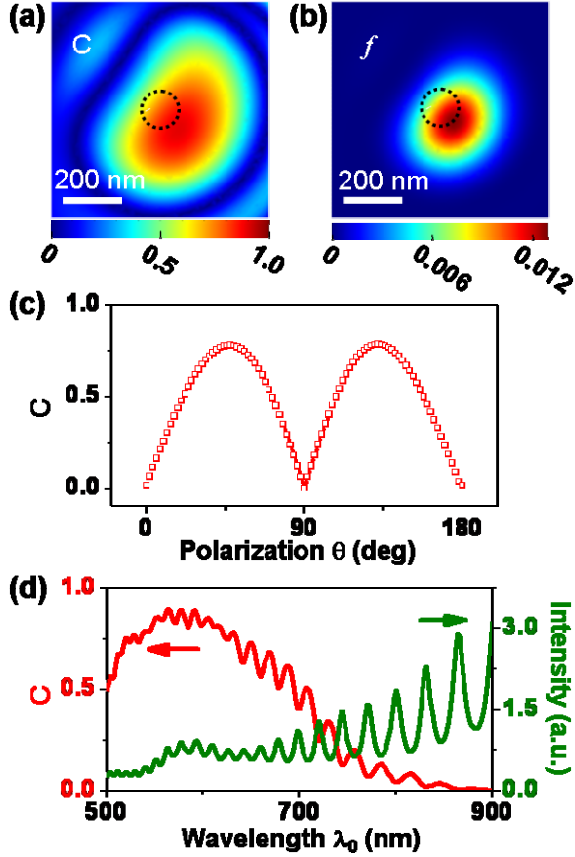

**FIG. 4** (a, b) Maps of degree of circular polarization C (a) and figure of merit $f$ (b) in a vertical plane 200 nm beyond the distal end of the nanowire. The black dotted circles indicate the cross section of the nanowire. (c) C at the center of (a) on the symmetric axis of the nanowire as a function of incident polarization angle θ. (d) C (red) vs. incident wavelength $\lambda_0$ at the same position as in (c). The transmission spectrum $I(\lambda_0)$ (green) is also shown. The incident wavelength is $\lambda_0$ = 632.8 nm in (a-c) and the polarization angle is $\theta = 45°$ in (a, b) and (d). The length of Ag nanowire is 5.0 μm and its radius is 60 nm.

This implementation of a broadband, subwavelength linear-to-circular polarization converter is highly desirable for numerous nanophotonics applications. Particularly, we envisage the realization of a metallic nanowire-based circularly polarized photon source as a unique new type of tip for scanning probe microscopies, e.g. scanning near-field optical microscopy [29] or tip-enhanced Raman spectroscopy. It is quite likely that single chiral molecule or 'artificial molecule'-light interactions may be facilitated by this source [2, 30]. Other possible applications include the study of spin dynamics of spintronic materials, the demonstration of new nanoplasmonic devices, [19-21] and subwavelength all-optical magnetic recording [31].

In conclusion, we have shown that chiral surface plasmon polaritons can be excited on metallic nanowires. This is accomplished by illumination of one end of the nanowire with linearly polarized light at a nominal 45 degree polarization angle with respect to the nanowire axis. Photons emitted from the end of the nanowire when excited in this manner can be highly circularly polarized, making this structure a broadband, subwavelength linear-to-circular polarization converter. This discovery of chiral electromagnetic surface waves creates new opportunities for the design of nanoscale integrated photonic components, and provides a subwavelength circular polarized light source that may be useful as a local probe of enantiomeric molecules and other reduced-symmetry nanoscale systems.

*hongxingxu@iphy.ac.cn